\documentclass[review,12pt]{elsarticle}

\usepackage{epsfig}
\usepackage{amssymb}
\usepackage{amsthm}
\usepackage{amsmath}
\usepackage{color}
\usepackage{bm}
\usepackage{amssymb}

\begin{document}

\begin{frontmatter}



\title{Electrical and mechanical properties of a fully hydrogenated two-dimensional polyaniline sheet}


\author[label1]{Meysam Bagheri Tagani\corref{cor1}}
 \ead{m{\_}bagheri@guilan.ac.ir}
\address[label1]{Department of Physics, Computational Nanophysics Laboratory (CNL),
University of Guilan, Po Box:41335-1914, Rasht, Iran.}

\begin{abstract}
Two-dimensional (2D) polyaniline sheet has been recently synthesized and showed that it is a semiconductor with indirect  band gap. In this research, we examine electrical and mechanical properties of a fully hydrogenated 2D polyaniline sheet $C_3NH$ using density functional theory. Results show that the $C_3NH$ sheet is an insulator with a band gap more than $5eV$. The sheet is quasi planner and dynamically stable confirmed by phonon band structure. Young modulus of the sheet is $275 N/m$. Ab-initio molecular dynamics simulations show that the $C_3NH$ sheet can be stable at $1000K$ indicating a high melting point. Tensile strain reduces the band gap of the sheet and electron effective mass. In return, hole effective mass is strongly dependent on the strain direction so that strain along zigzag (armchair) increases (reduces) hole effective mass. our findings show that $C_3NH$ sheet is a promising candidate for electronic and optoelectronic applications and strain engineering can be used to tune its properties.
\end{abstract}

\begin{keyword}
Polyaniline sheet, Density functional theory, Young modulus, Phonon modes

\end{keyword}

\end{frontmatter}


\section{Introduction}
\label{Intro}
\par Nowadays, synthesis and characterization of two-dimensional materials (2D) has attracted a lot of attention. The story of 2D materials begun after discovery of graphene at 2004 \cite{graphene}. Graphene as a monolayer of carbon atoms has individual and different properties in comparison with prior materials. High carrier mobility and thermal conductivity, excellent mechanical ability, quantum hall effect, and linear band structure around Fermi level make it a fascinating material for scientists \cite {graphene1, graphene2, graphene3, graphene4}. An atom thickness of graphene introduces it as a promising candidate for next-generation field effect transistor, however, absence of band gap is a great challenge in this field. Two strategies have been considered to solve the problem of band gap: creating a band gap in graphene structure or finding other 2D materials with intrinsic band gap. The former led to graphene nanoribbons \cite{GNR1, GNR2}, graphene antidote lattices \cite{GAL1, GAL2}, and mutilayer graphene sheets \cite{MGS}. The later added silicene \cite{silicene}, germanene \cite{germanene}, phosphorene \cite{phosphorene}, stanene \cite{stanene}, and borophene \cite{borophene} to 2D material family which some of them are semiconductor. In recent years a lot attention has been paid to 2D materials to investigate electrical, mechanical, and magnetic properties of them \cite{datta1, datta2, datta3, datta4, bagheri1, bagheri2, bagheri3}.

\par Mahmood et al. has recently synthesized a new 2D planner material from carbonized organic single crystal \cite{mahmood1}. Two-dimensional polyaniline sheet with empirical formula of $C_3N$ is an indirect semiconductor which has attracted a lot of attention because of its similarity with graphene. Indeed, $C_3N$ can be considered as an improved version of nitrogenated holey two-dimensional structure with $C_2N$ stoichiometry synthesized at 2015 \cite{mahmood2}. Yang and coworkers has prepared $C_3N$ sheet by polymerization of 2,3- diaminophenazine and showed that field-effect transistors made from $C_3N$ display an on-off current ratio of $5.5\times 10^{10}$ \cite{Yang}. Xu et al. showed that $C_3N$ sheet is an excellent anode material for lithium-ion batteries\cite{Xu}. $C_3N$ sheets have also attracted attention from theoretical point of view \cite{makaremi, cui, zhou, wang, hong}. It was shown that $C_3N$ can be an excellent sensor for probing toxic gases like $NO_2$ and $SO_2$ \cite{cui}. Optical and mechanical properties of $C_3N$ has been investigated in Ref.\cite{zhou}. Bilayer $C_3N$ shows anisotropic carrier mobility \cite{wang} and high thermal conductivity \cite{hong}. We studied $C_3N$ nanoribbons and found that some of them can be half-metal \cite{Bagheri}.

\par Hydrogenation of single element 2D sheets can significantly change their electronic, mechanical and optical properties. Theoretical predictions showed that graphane, a fully hydrogenated graphene sheet, unlike graphene is a semiconductor \cite{sofo}. Later, Elias et al. synthesized graphane and confirmed the theoretical predictions \cite{elias}. In this research, we study a fully hydrogenated 2D polyaniline sheet ($C_3NH$) with formula $C_3NH_3$ using ab-initio calculations. Our investigation shows that just carbon atoms can host hydrogen and adding hydrogen atoms to nitrogen results in deformation of the structure. Phonon band structure shows the $C_3NH$ is a stable and quasi 2D sheet. The electronic band gap of the sheet is about $5eV$ and the sheet can be stable under uniaxial tensile strain as high as $10\%$. Ab-initio molecular dynamics simulations show that the structure is stable at $1000K$.

\section{\label{sec:level2}Simulation Details}
\par In this paper we have used SIESTA package \cite{siesta}, which is based on density functional theory (DFT) and uses strictly localized basis sets, to study electronic properties of $C_3NH$ sheet. Exchange-correlation functionals are described using GGA-PBE approximation and ultra-soft pseudopotentials are used for core electrons. Brillouin zone is meshed by $51\times 31\times 1$ K point sampling. Double zeta single polarized (DZP) basis set is chosen to describe valance electrons. All studied structures are carefully relaxed so that the maximum force which is applied on each atom is less than $0.001 eV/$\AA ~and stress on the cell is less than $0.001 GPa$. Cutoff energy is set to $75 Ha$ and Van der Walls correction as Grimme DFT-D2 \cite{grimme} is considered to describe interaction between hydrogen and carbon atoms. A $20$\AA ~ vacuum is set perpendicular to the sheet to prevent artifact interactions between the sheet and its image.

\par To investigate the mechanical properties of the sheet a supercell composed of 28 atoms is considered. Dynamical stability of the sheet is studied using phonon band structure for each strain. For phonon calculations, we have used a supercell composed of 420 atoms. Phonon calculation are very time consuming and needs huge CPUs, therefore, we have used DFT-tight binding method (DFTB) on the basis of Slater-Koster parameters \cite {SK} which is a very fast method and executable on laptop. The first Brillouin zone of the supercell is meshed by a $8\times 8\times 1$ k-point sampling and cutoff density is $40 Ha$. Comparison between the phonon band structure obtained from DFTB and DFT for pristine $C_3N$ shows that DFTB provides reliable results for phonon structure in $C_3N$ and its derivatives.
\par Thermal stability of the structure is investigated using ab-initio molecular dynamics (AIMD). A $4\times 4\times 1$ supercell is considered and Brillouin zone is meshed by a $3 \times 3 \times 1$ K-point sampling. The simulation is carried out in NVT ensemble with Nose-Hoover thermostat approach. The simulation is preformed for $3ps$ with time step of $2fs$. Thermal stability of the sheet is studied at $300,~500,$~ and $1000K$.

\begin{figure}[htb]
\centering
  \includegraphics[height=100mm,width=150mm,angle=0]{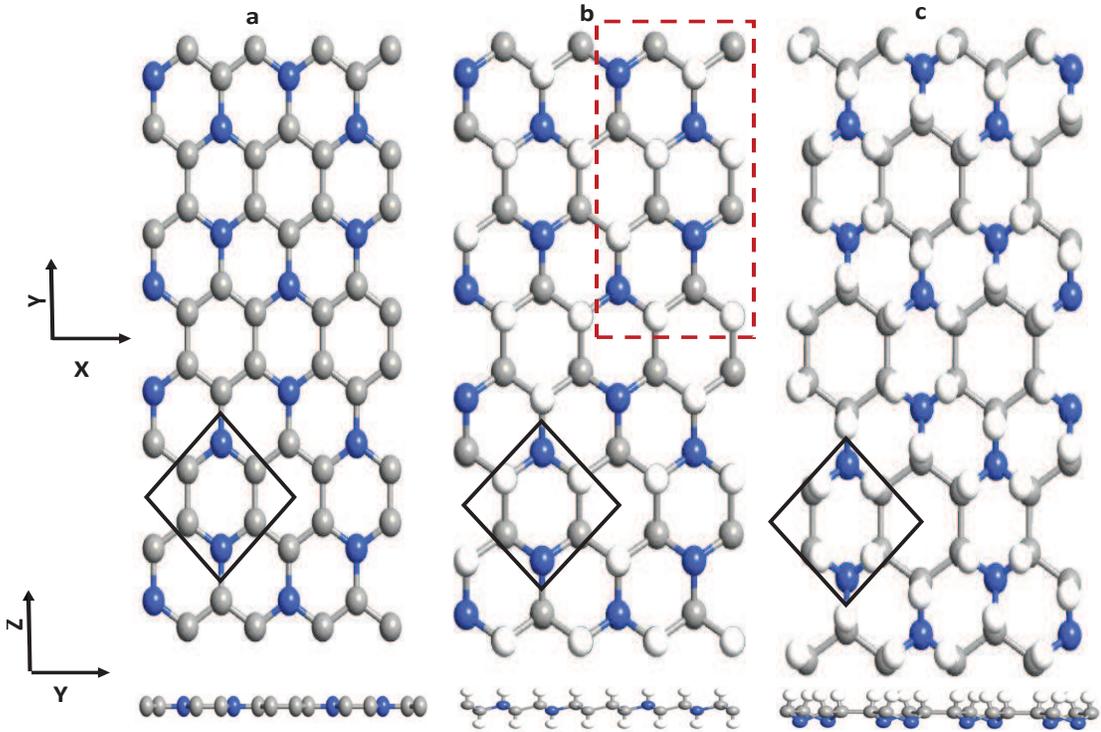}
  \caption{Top and side view of (a) $C_3N$ sheet, (b) chair, and (c) boat like $C_3NH$ sheet. Unit cell is shown by rhombus. Conventional unit cell considered for strain calculations s shown by rectangular. Carbon, nitrogen and hydrogen atoms are shoen by gray, blue, and whit balls, respectively. }
  \label{Fig1}
\end{figure}

\section{\label{sec:level3}Results and Discussions}
\label{Numerical results}
\par Fig. 1 shows the structure of a polyaniline monolayer, $C_3N$, and two different hydrogenated polyaniline monolayer, $C_3NH$ with hexagonal symmetry. Our investigation shows that just carbon atoms are enable to absorb hydrogen (see Fig. S1) and the structure is still quasi planner. Adding hydrogen atoms to nitrogen ones having three electrons for bonding disturbs the structure because of breaking C-N bonding. It is clear in Fig. S1 that there is no bonding between nitrogen and hydrogen and the distance between them is larger than $1.9$ \AA~. We consider two different scenarios for hydrogenating the structure like graphane: hydrogen atoms are located at two different $z$ planes as chair-like, Fig. 1b, or they are at the same plane as boat-like, Fig. 1c. To compare the stability of two different $C_3NH$ structures the formation and binding energies of the structures are calculated as follows:
\begin{equation}
E_f=E_T(sheet)-\frac{1}{2}(N_N E(N_2)+N_HE(H_2))-N_CE({C_g})
\end{equation}
\begin{equation}
E_b=E_T(sheet)-(N_N E(N)+N_HE(H))-N_CE(C)
\end{equation}
where $N_N (N_H)$ is the number of nitrogen (hydrogen) atoms in the unit cell, and $E(N_2) (E(H_2))$ is total energy of nitrogen (hydrogen) molecule. $N_C$ denotes the number of carbon atoms, and $E(C_g)$ is energy of each carbon atom in graphite. For binding energy calculation we have used ground state energy of isolated atoms obtained from spin-dependent calculations. Binding energy of chair-like structure is $-5.473 eV/atom$, while it is $-5.128 eV/atom$ for boat-like structure. Formation energy of the chair-like structure is equal to $-2.45 eV$, while it is equal to $2.38 eV$ for the other structure. Therefore, we conclude that the structure with hydrogen atoms in opposite sides is energetically more favorable and we start to analyze this structure. Same trend is observed for graphane case but with lower energy difference between two different configurations \cite{sofo}.
\begin{figure}[h]
\centering
  \includegraphics[height=100mm,width=150mm,angle=0]{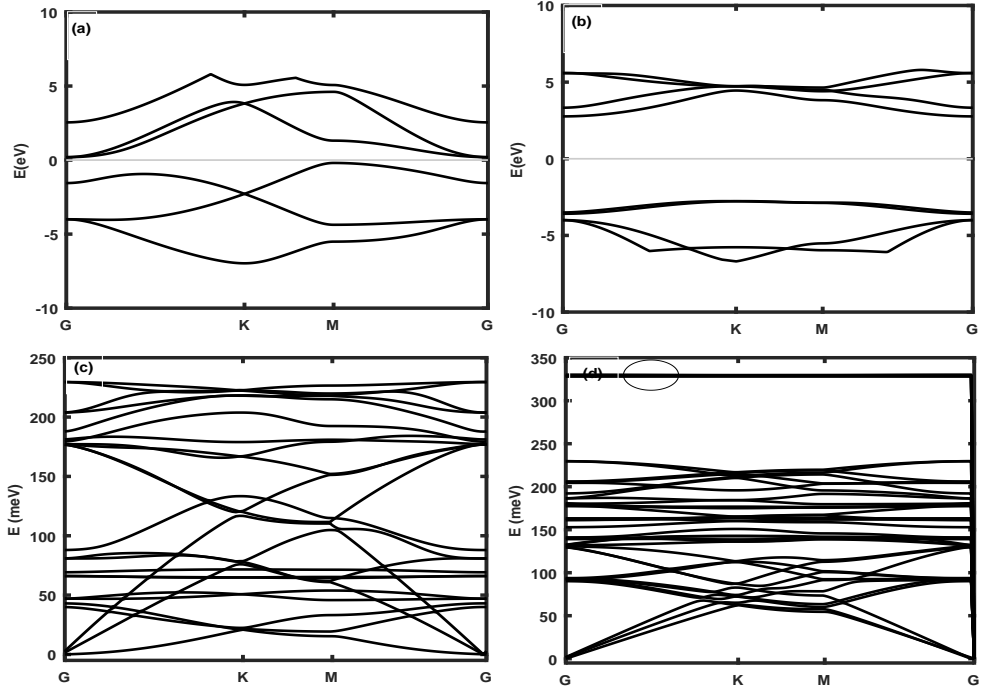}
  \caption{(a), and (c) electron and phonon band structure of $C_3N$ sheet. (b) and (d) electron and phonon band structure of $C_3NH$ sheet, respectively.}
  \label{Fig2}
\end{figure}
\par $C_3N$ has a hexagonal lattice with lattice constant of $4.87$ \AA~ and is composed of six carbon and two nitrogen atoms bonded together with $sp^2$ hybridization. C-N bond length is equal to $1.41$ \AA,~ and C-N bond length is $1.40$ \AA. These results are in consistent with perviously experimental and theoretical reports \cite{Yang, zhou}. C3NH shows a hexagonal symmetry with lattice constant of $4.90$ \AA. Adding hydrogen does not disturb the lattice symmetry and increase of lattice constant is ignorable. However, bond length, angles and atom positions are significantly altered after hydrogenation. A transition from $sp^2$ to $sp^3$ is observed resulting in the increase of C-C and C-N bond length. C-C bond length is equal to $1.53$ \AA~ i.e. 8 percent increase, and C-N is equal to $1.46$ \AA~ i.e. 4 percent increase. C-H bond length is equal to $1.12$ \AA. The angle of C-N-C is equal to $110^0$ indicating $sp^3$ hybridization.
The obtained C-C bond length is equal to the graphane case \cite{sofo}. The buckling between N atoms is equal to $0.45$ \AA~ and to $0.50$ \AA~ for carbon ones.
\begin{figure}[h]
\centering
  \includegraphics[height=100mm,width=150mm,angle=0]{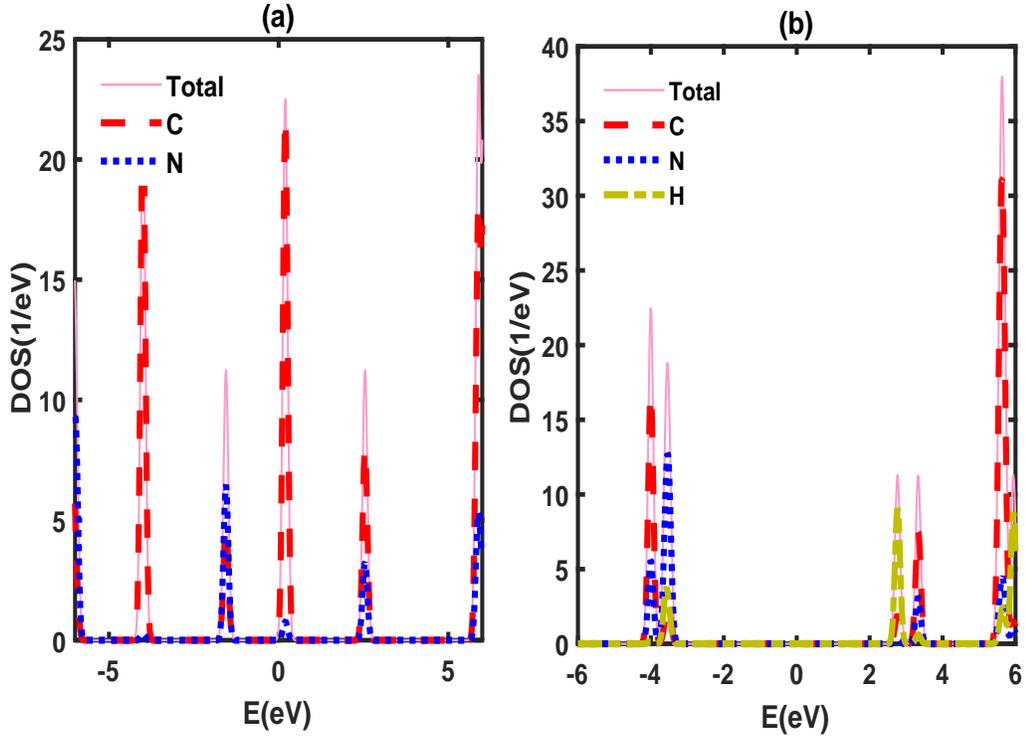}
  \caption{Density of states for (a) $C_3N$ and (b) $C_3NH$ sheet.}
  \label{Fig3}
\end{figure}

\par Electron and phonon band structure of $C_3N$ and $C_3NH$ is plotted in Fig. 2. $C_3N$ is a semiconductor with indirect band gap of $0.39 eV$. Top of valance band is located at $M$ point and bottom of conduction band is located at $\Gamma$ point. The calculated band gap is consistent with perviously theoretical and experimental reports \cite{mahmood1, Yang, zhou}. Nitrogen and carbon atoms have equal participation at the top of valance band, but the bottom of conduction band is belonging to carbon atoms as shown in Fig. 3a. Our results reveal that $C_3NH$ is an insulator with band gap of $5.53 eV$ which is $2 eV$ more than graphane case \cite{sofo}. It is well-known that GGA-PBE functional underestimates the band gap, so we expect that calculations performed by hybrid functional like HSE06 or many-body perturbation theory like GW predict larger band gap.  Maximum of valance band is located at $\Gamma-K$ direction and near $K$ point. Hydrogenation strongly reduces the dispersion of valance band so electrons are completely localized. In addition, the Dirac cone observed in $C_3N$ sheet at $E=-2 eV$ is disappeared after hydrogenation. It is found that the nitrogen atoms have the dominant role at the creation of maximum of valance band, while the minimum of conduction band is related to hydrogen atoms as seen in Fig. 3b. Appearance of buckling, change of hybridization from $sp^2$ to $sp^3$, and increase of bond length between atoms are the main factors in the increase of band gap and localization of electrons in valance band. Dynamical stability of the structures is investigated by computing phonon band structure. Both structures are stable because there is no imaginary phonon mode. The phonon band structure obtained for $C_3N$ in our work is similar to the one presented for the same structure based on DFT calculations, therefore, DFTB method used in this work can be considered as a suitable choice for computing phonon modes. Comparison between DFTB and DFT based phonon calculations are presented in Fig. S2 which shows reliable accuracy of DFTB method. Acoustic flexural mode with quadratic dispersion around $\Gamma$ point is observed in $C_3N$ sheet and crosses with other acoustic modes at $K$ point. This mode is a characteristic of low dimensional crystals and indicates the motion of atoms toward out of the plane. This mode has a low energy because there are no atoms out of the plane. Several phonon bands cross each other at $K$ point which is a property of hexagonal structures \cite{sahin1, sahin2}.
\begin{figure}[h]
\centering
  \includegraphics[height=100mm,width=150mm,angle=0]{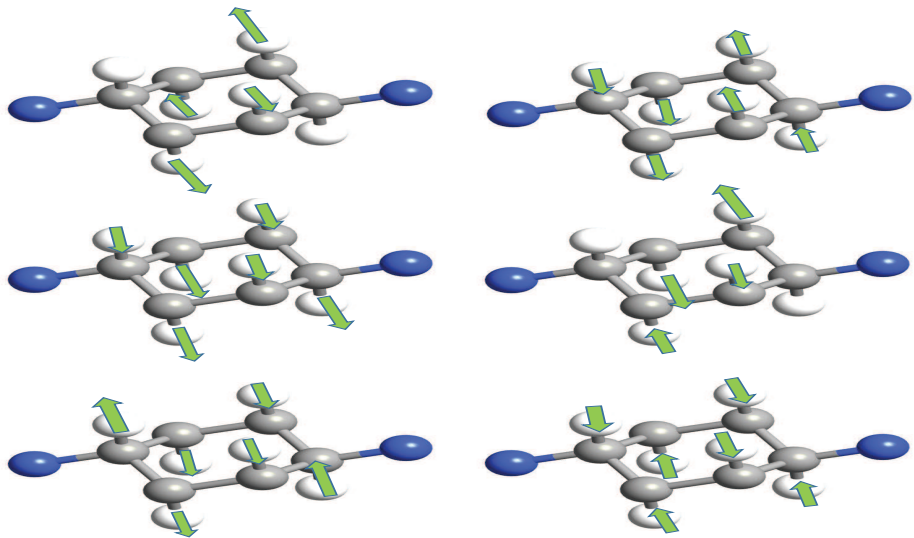}
  \caption{Displacement of atoms in $C_3NH$ sheet for phonon modes shown by circle in Fig. 2d.}
  \label{Fig4}
\end{figure}
\begin{figure}[h]
\centering
  \includegraphics[height=80mm,width=80mm,angle=0]{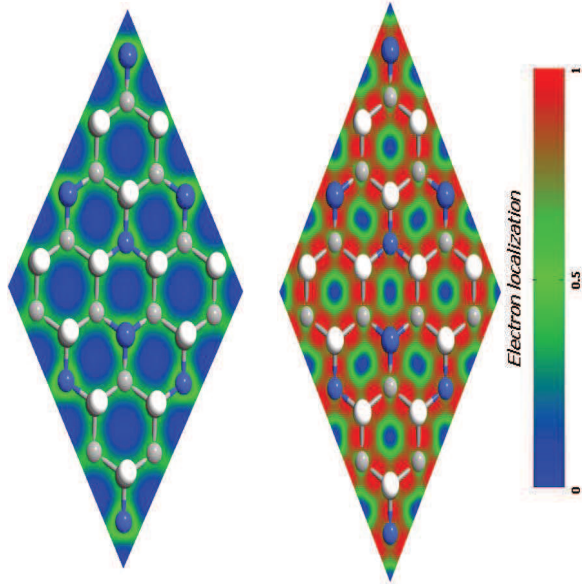}
  \caption{Electron density (left panel), and electron localization function (right panel).}
  \label{Fig5}
\end{figure}
The flexural mode becomes linear like transverse and longitudinal acoustic modes in $C_3NH$ because there are hydrogen atoms and some buckling in \textit{z} direction. There are six degenerate phonon bands in energies more than $300 meV$, shown by circle, which are absent in $C_3N$ sheet. These modes are related to the motion of hydrogen atoms under condition that all carbon and nitrogen atoms are fixed. The different modes indicating the motion of hydrogen atoms are plotted in Fig. 4. To gain more insight about the bonding in $C3NH$, electron density and electron localization function (ELF) are presented in Fig.5. It is clear that the electrons are distributed between carbon and nitrogen atoms showing a covalent bonding. However, the electron density is more on nitrogen atoms because they are more electronegative than carbon ones. In addition, ELF map confirms the covalent bonding because electrons are distributed all over the sheet.

\par To investigate the mechanical properties of $C_3NH$ sheet a rectangular conventional cell composed of 28 atoms is considered as shown in Fig. 1b. The cell is a rectangular and we study its response to strain along \textit{x} direction, zigzag direction, and \textit{y} direction, armchair direction. Young's modulus is computed by applying strain ($\epsilon$) in linear response regime, $0<\epsilon<+2\%$, with a step of $0.0025\%$ and measuring slope of stress-strain relation. The strain is defined as $\epsilon=\frac{L-L_0}{L_0}\times100$ where $L_0(L)$ is original (stretched) lattice constant. To compute the stress, we apply tensile strain in one direction and allow atoms' position and the lattice constant in another direction to relax so that the force acting on each atom and stress on the cell become less than $0.001 eV/\AA$ and $0.001 GPa$, respectively.Compressive strain isn't considered because it can result in wrinkle of 2D materials.
\begin{figure}[h]
\centering
  \includegraphics[height=100mm,width=150mm,angle=0]{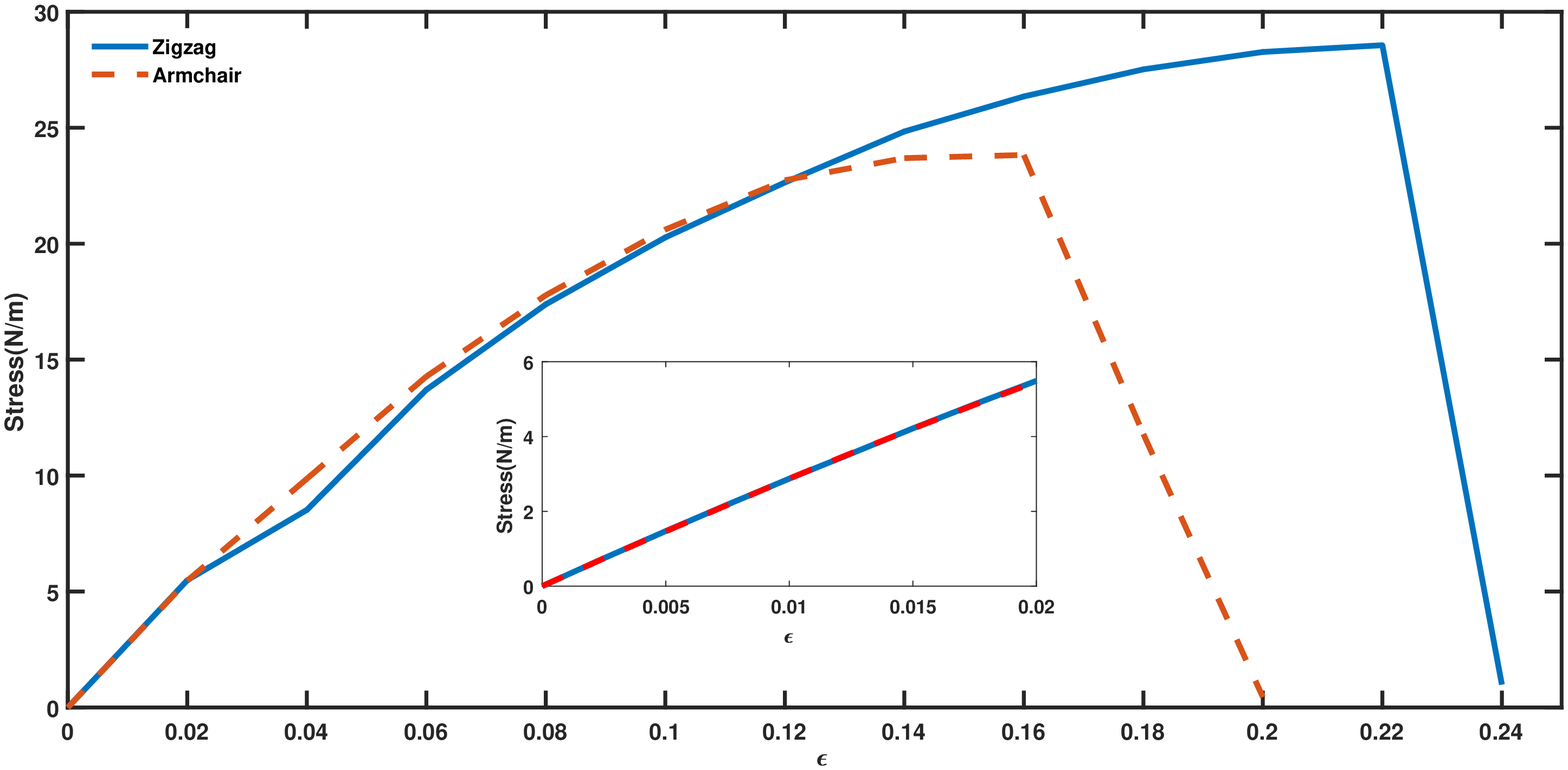}
  \caption{Stress-strain relation. Inset shows linear regression of stress-strain.}
  \label{Fig6}
 \end{figure}
 \begin{figure}[h]
\centering
  \includegraphics[height=100mm,width=150mm,angle=0]{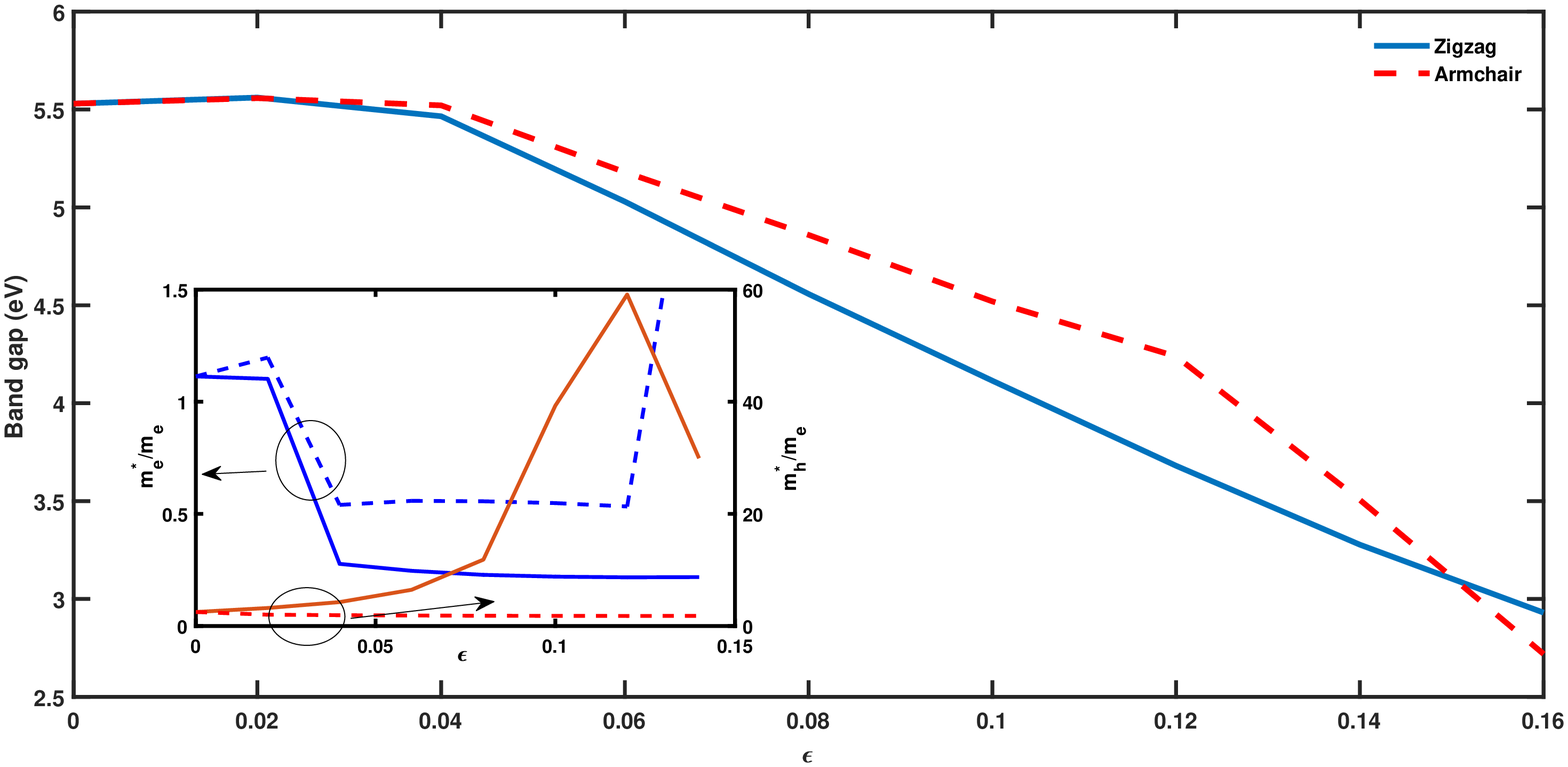}
  \caption{Variation of band gap under tensile strain. Inset shows electron and hole effective mass at minimum of conduction band and maximum of valance band as a function of tensile strain along zigzag (solid line) and armchair (dashed line) direction.}
  \label{Fig7}
 \end{figure}

\par Fig. 6 shows the strain-stress relation under uniaxial tensile strain along zigzag and armchair direction. The Young's modulus is obtained by fitting stress-strain curves based on linear regression up to $2\%$ along zigzag and armchair directions as shown in the inset. The Young's modulus is equal to $E_Z=274.5 N/m$ along zigzag direction and $E_A=273.7 N/m$ along armchair direction. The Young's modulus of a pristine $C_3N$ sheet was reported to be $E=355.6 N/m$ \cite{zhou} showing the hydrogenation reduces the mechanical ability of the sheet up to $23\%$. The Young's modulus of graphene is $342.2 N/m$ \cite{Liu}, while it is equal to $243 N/m$ for graphane \cite{cad} i.e. a $29\%$ reduction after hydrogenation. Therefore, it is concluded the hydrogenation process has less negative effect on mechanical properties of $C_3N$ sheet in comparison to graphene case. Critical strain is equal to $16\%$ and $22\%$ for armchair and zigzag directions, respectively, and corresponding ideal strengths are $23.8 N/m$, and $28.56 N/m$. The calculated Young modulus for $C_3NH$ is higher than $\chi_3$-borophene ($180N/m$)\cite{B1}, $\beta_{12}$-borophene ($207N/m$)\cite{B1}, and $MoS_2$ sheet ($200N/m$)\cite{M1} which were experimentally synthesized.
\begin{figure}[h]
\centering
  \includegraphics[height=100mm,width=150mm,angle=0]{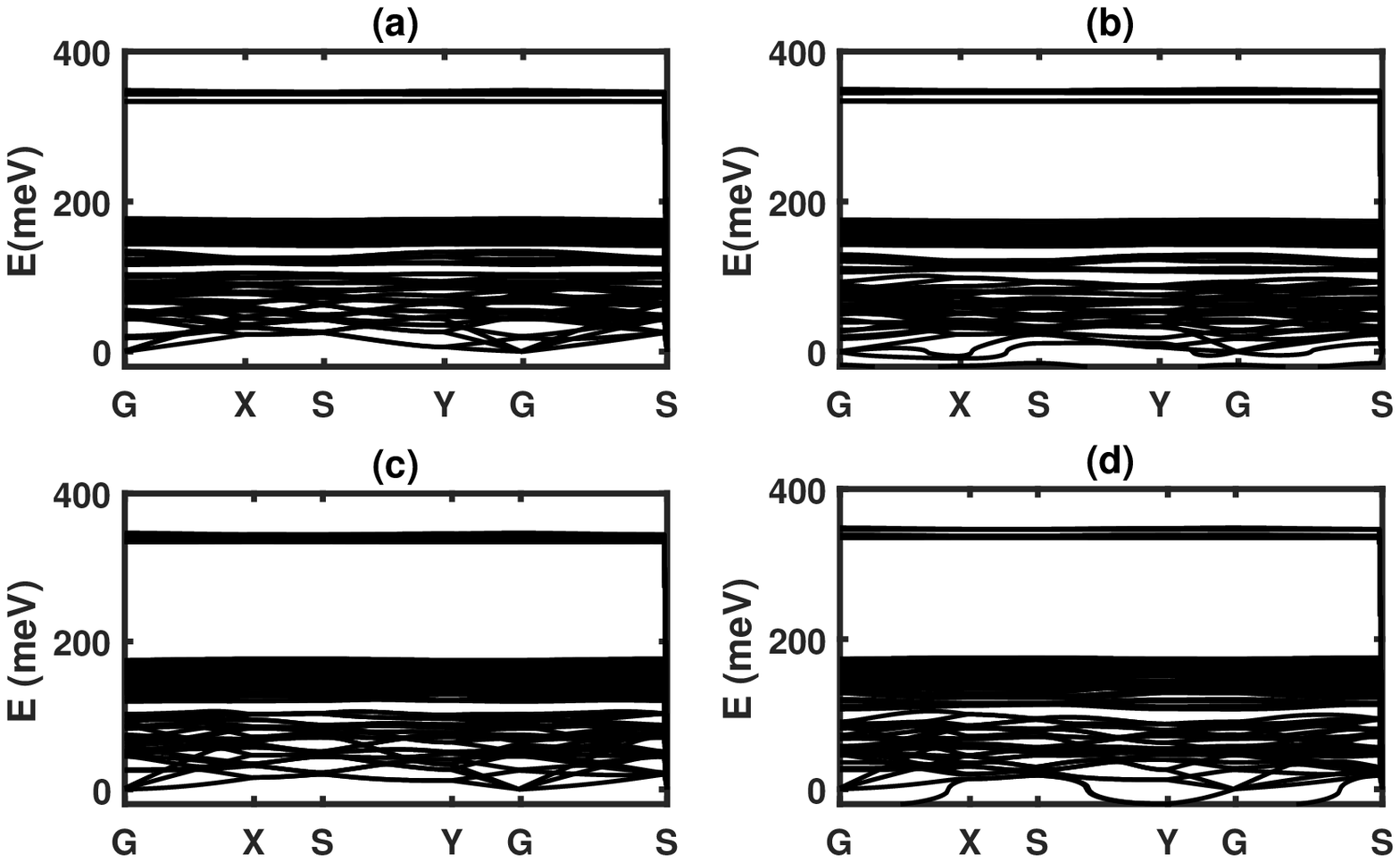}
 \caption{Phonon band structure of $C_3NH$ sheet under tensile strain along (a), and (b) zigzag direction and $\epsilon=12 \%$~ and $14\%$, and (c), (d) armchair direction and $\epsilon=8 \%$~ and $10\%$, respectively.}
  \label{Fig8}
 \end{figure}

\par Strain has a significant influence on the electrical properties of two-dimensional materials so that graphene becomes a semiconductor under external strain. Therefore, we investigate the variation of band gap of $C_3NH$ sheet under uniaxial tensile strain in Fig. 7. Results show that the strain can reduce the band gap. The reduction is dependent on the strain direction and its magnitude as the variation is more intense for strain along zigzag direction. It is interesting that the strain reduces the gap in $C_3NH$, therefore, strain engineering can be used to transform $C_3NH$ from an insulator to a semiconductor which is suitable for optoelectronics and field effect transistors. Variation of electron and hole effective mass as a function of tensile stress is shown in the inset. Effective mass is inversely dependent on $\frac{\partial^2{E_k}}{\partial{k^2}}$. Results show that the electron becomes lighter with increase of strain. On the other hand, hole effective mass is dependent on the strain direction. The hole becomes heavier when strain is applied along armchair direction, while it looks lighter for strain in zigzag direction. Strain can be used as a tool to tune transport properties of the sheet.  Fig. S4 and S5 shows the band structure of pristine and stretched $C_3NH$ under strain in different directions. It is clear that the strain lowers the minimum of conduction band and raises the maximum of valance band. In addition, we have calculated electron and hole effective mass versus strain presented in table S1 and S2.
\begin{figure}[h]
\centering
  \includegraphics[height=100mm,width=150mm,angle=0]{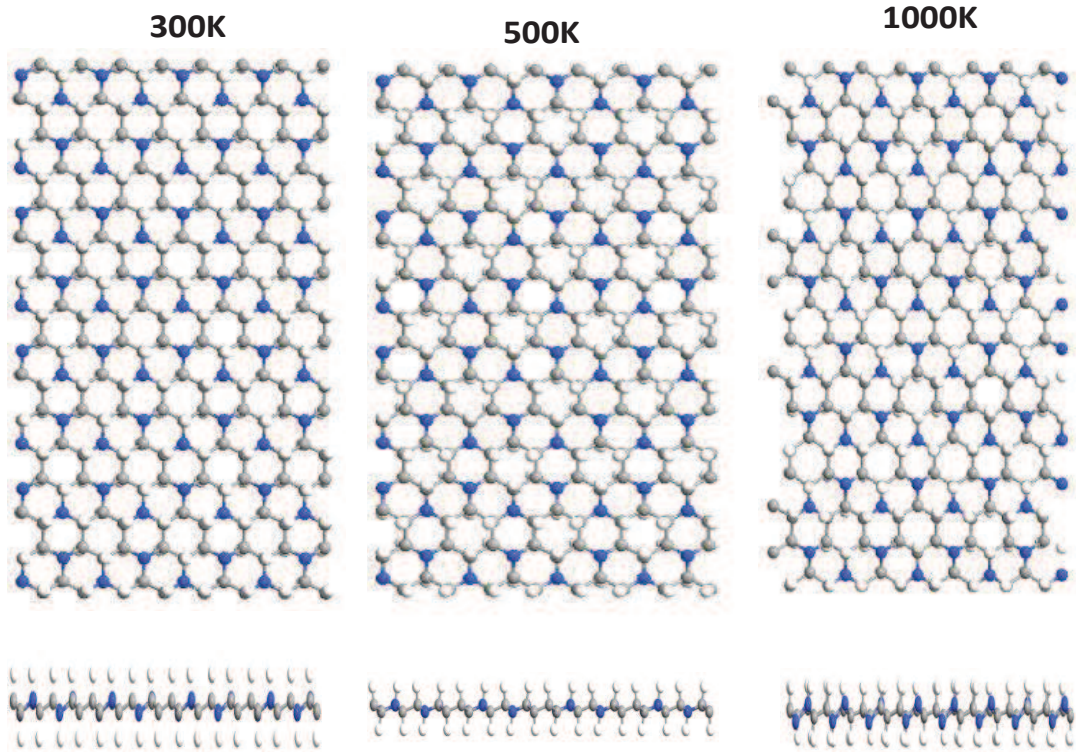}
 \caption{AIMD simulation of $C_3NH$ sheet in three different temperatures. It is clear that the sheet is stable at $1000K$.}
  \label{Fig9}
 \end{figure}
\par Dynamical stability of the structure under tensile strain is an important challenge which can be investigated using phonon band structure. Fig. 8 shows the phonon band structure of $C_3NH$ under different strengths and orientations of tensile strain. It is revealed that the sheet can be stable up to $\epsilon=12\%$ in zigzag direction and after that, the sheet will be unstable. Instability is confirmed by imaginary phonon modes as shown in Fig. 8b. Stability of the sheet is weaker for tensile strain along armchair direction so that the sheet is unstable for $\epsilon>10\%$. Results show that $C_3NH$ structure has a mechanical anisotropy so that it is stiffer in zigzag direction than armchair one. In addition, the structure is stable under tensile strain less than $\epsilon<10\%$. Our results are based on DFTB, therefore, we expect that measured stability will be more than our prediction if DFT approach is used. Moreover, substrate can increase mechanical ability of the $C_3NH$ sheet.
\par Fig.9 shows the AIMD simulation results at $300, 500,$~ and $1000 K$. As it is clear from the Fig, the sheet is stable at temperature as high as $1000K$ indicating its high melting point. Its high stability comes from strong C-C and C-N bondings. Increase of temperature brings some distortion in the structure so that the buckling is increased. However, the sheet saves hexagonal rings and stays quasi planner.  Techniques  employed to hydrogenate carbon nanotubes and graphene can be used to synthesize $C_3NH$ sheet. For example, to prepare $C_3NH$ sheet, hydrogen atoms can be chemisorbed on carbon atoms of polyaniline sheet by ultra-violet\cite{uv} or X-ray spectroscopy \cite{xr}. In addition, $C_3N$   sheet can be exposed to a cold hydrogen plasma for several hours to form $C_3NH$

\section{\label{sec:level4}Conclusions}
We have studied electrical and mechanical properties of a fully hydrogenated two-dimensional polyaniline sheet ($C_3NH$) using density functional theory. Two different configurations as chair-like and boat-like $C_3NH$ were considered. We found that chair-like configuration is more favorable than the other from energy point of view. From phonon band structure, it is found that both structure are dynamically stable. The $C_3NH$ sheet is an insulator with band gap of $5.5 eV$. Young modulus of the sheet is $275 N/m$ and uniaxial tensile strain can reduce the band gap of the sheet. Thermal stability of the sheet is studied using ab-initio molecular dynamics simulations and found that the melting point of the sheet is higher than $1000 K$. Electron and hole effective mass versus strain is also investigated and  realized that electrons (holes) become lighter (heavier) under tensile strain.

\section*{Conflicts of interest}
There are no conflicts to declare.

\section{Acknowledgment}

\end{document}